\documentclass[letterpaper]{article}

\newcommand{\citep}[1]{\cite{#1}}

\usepackage {amssymb}
\usepackage [dvips] {graphicx}
\usepackage {epsfig}
\usepackage [tbtags] {amsmath}
\usepackage {amsfonts}
\usepackage {mathrsfs}
\usepackage {psfrag}
\usepackage {pstricks}
\usepackage {float}
\usepackage {rotating}
\usepackage {hyphenat}
\usepackage [mathcal] {eucal}

\begin{document}

%
%
\newcommand{\bsdd}{\boldsymbol{D}}
\newcommand{\bsf}{\boldsymbol{\mathrm{f}}}
\newcommand{\bsff}{\boldsymbol{\mathrm{F}}}
\newcommand{\bsmm}{\boldsymbol{\mathrm{M}}}
\newcommand{\bsd}{\boldsymbol{\mathrm{d}}}
\newcommand{\bsi}{\boldsymbol{i}}
\newcommand{\bsj}{\boldsymbol{j}}
\newcommand{\tn}{\tilde{n}}
\newcommand{\bsn}{\boldsymbol{n}}
\newcommand{\tbsn}{\boldsymbol{\tilde{n}}}
\newcommand{\bst}{\boldsymbol{t}}
\newcommand{\bsu}{\boldsymbol{u}}
\newcommand{\bsw}{\boldsymbol{w}}
\newcommand{\bsx}{\boldsymbol{x}}
\newcommand{\bsy}{\boldsymbol{p}}
\newcommand{\bsz}{\boldsymbol{q}}
\newcommand{\bsone}{\boldsymbol{1}}
\renewcommand{\epsilon}{\varepsilon}
\newcommand{\bsepsilon}{\boldsymbol{\epsilon}}
\newcommand{\bsnabla}{\boldsymbol{\nabla}}
\newcommand{\bssigma}{\boldsymbol{\sigma}}
\newcommand{\ca}{C_{a}}
\newcommand{\ceq}{C_{v}^{eq}}
\newcommand{\ceqo}{C_{v_{0}}^{eq}}
\newcommand{\conc}{C_{v}}
\newcommand{\cs}{C_{s}}
\newcommand{\elasc}{\mathbb{C}}
\newcommand{\msb}{\mathscr{B}}
\newcommand{\msf}{\Gamma}
\newcommand{\mcg}{\mathcal{G}}
\newcommand{\evf}{e_{v}^{f}}
\newcommand{\evd}{e_{v}^{d}}
\newcommand{\egb}{e_{m}}
\newcommand{\muv}{\mu_{v}}
\newcommand{\muvo}{\mu_{v}^{eq}}
\newcommand{\rmd}{\mathrm{d}}
\newcommand{\rmda}{\mathrm{dV}}
\newcommand{\rmdl}{\mathrm{dS}}
\newcommand{\bsphi}{\boldsymbol{\Phi}}
\newcommand\etal{\textit{et al.}}
%
%
\renewcommand{\bar}[1]{\overline{#1}}
\newcommand{\order}[1]{\mathcal{O}(#1)}
\newcommand{\dssum}{\displaystyle\sum}
%
%
%
%
\makeatletter%
\def\@captionsize{\normalsize}
\def\@overcaptionskip{8\p@}
\intextsep 8\p@ \@plus 4\p@ \@minus 2\p@
\renewcommand\floatc@plain[2]{\setbox\@tempboxa\hbox{\@captionsize #1: #2}%
  \ifdim\wd\@tempboxa>\hsize \@captionsize #1: #2\par
  \else\hbox to\hsize{\hfil\box\@tempboxa\hfil}\fi}  
\renewcommand\fs@plain{\def\@fs@cfont{\rmfamily}\let\@fs@capt\floatc@plain
  \def\@fs@pre{\vskip \intextsep}\def\@fs@post{\vskip \intextsep}\def\@fs@mid{\vskip \@overcaptionskip}%
  \let\@fs@iftopcapt\iftrue} 
\renewcommand\@makefntext[1]{%
    \parindent 1em%
    \noindent
    \hb@xt@1.0ex{\hss\@makefnmark}#1}
\makeatother%
\newfloat{algorithm}{thp}{loa}
\floatname{algorithm}{Box}
\newcounter{Remctr}
\newcommand{\remark}{\stepcounter{Remctr}\noindent\textbf{Remark \arabic{Remctr}}~~}
\setlength{\parindent}{0pt}
\setlength{\parskip}{1.2ex plus 1.2ex minus 0.4ex}
\addtolength{\skip\footins}{1.8em}
\renewcommand{\thefootnote}{\fnsymbol{footnote}}
\newcommand{\sm}{\small}

\title{Advances in the numerical treatment of\\ grain-boundary migration: Coupling with\\ mass transport and mechanics}

\author{Hashem M. Mourad${}^{\dagger}$, Krishna Garikipati${}^{\ast,\ddagger}$}
\date{\textit{\small Department of Mechanical Engineering, University of Michigan\\ Ann Arbor, Michigan 48109, USA \normalsize}}
\stepcounter{footnote}
\footnotetext{Corresponding author.}
\stepcounter{footnote}
\footnotetext{Present address:~~Theoretical Division, T-03, Los Alamos National Laboratory, Los Alamos, New Mexico 87545, USA.}
\stepcounter{footnote}
\footnotetext{Email:~~Krishna Garikipati $<$\texttt{krishna@umich.edu}$>$}
\maketitle

\begin{abstract}
This work is based upon a coupled, lattice-based continuum formulation that was previously applied to problems involving strong coupling between mechanics and mass transport; e.g.\ diffusional creep and electromigration~\citep{Garikipatietal:01,GarikipatiBassman:01}. Here we discuss an enhancement of this formulation to account for migrating grain boundaries. The level set method is used to model grain-boundary migration in an Eulerian framework where a grain boundary is represented as the zero level set of an evolving higher-dimensional function. This approach can easily be generalized to model other problems involving migrating interfaces; e.g.\ void evolution and free-surface morphology evolution. The level-set equation is recast in a remarkably simple form which obviates the need for spatial stabilization techniques. This simplified level-set formulation makes use of velocity extension and field re-initialization techniques. In addition, a least-squares smoothing technique is used to compute the local curvature of a grain boundary directly from the level-set field without resorting to higher-order interpolation. A notable feature is that the coupling between mass transport, mechanics and grain-boundary migration is fully accounted for. The complexities associated with this coupling are highlighted and the operator-split algorithm used to solve the coupled equations is described.

\vspace{2ex}

\end{abstract}

%
%
%
%
%
%
\section{Introduction} \label{sec:intro}

In a recent paper, \mbox{Garikipati \etal\ \citep{Garikipatietal:01}} presented a coupled continuum field formulation for the interaction of electric effects, mechanical response and self-diffusion. Their approach drew upon earlier work by Larch\'e and Cahn~\citep{LarcheCahn:73,LarcheCahn:85}, Nix and co-workers~\citep{Nix:81,HirthNix:85}, G\'enin~\citep{Genin2:95}, Bower and Freund~\citep{BowerFreund:95}, \mbox{Xia \etal\ \citep{Xiaetal:97}}---to name but a few. In Reference~\citep{Garikipatietal:01}, a review of these works and others was presented, computational techniques were developed based on the finite element method, and several initial and boundary value problems involving diffusional creep (Nabarro-Herring and Coble creep) were solved.

The formulation introduced in~\citep{Garikipatietal:01} was extended to interdiffusion, with dopants in silicon as motivation, by Garikipati and Bassman~\citep{GarikipatiBassman:01}. In the current paper, we present a further extension of that same formulation to account for the interaction of grain-boundary migration with stress-driven self-diffusion and electromigration in polycrystalline solids. 

Different strategies used to model grain-boundary motion in a computational setting are described in the literature. Sun and Suo~\citep{Sun:97a,Sun:97b} developed a two-dimensional finite element formulation, based on the idea that the energy dissipated during the motion of the boundary must equal the reduction in the free energy of the system. This formulation was used to study several problems including grain growth in a thin film and the competition between surface grooving and grain-boundary migration. A similar variational formulation was developed by Cocks and Gill~\citep{Cocks:96,Gill:96} and used to model the evolution of a large network of grains in two dimensions. Another model was developed by \mbox{Zhao \etal\ \citep{Zhao:96}}, based on the variational formulation of Reitich and Soner~\citep{Reitich:96} and using the level set method of Osher and Sethian~\citep{Osher:88}.

The level set method is also used in the present paper as an interface-capturing technique. However, the goal of the current work is not merely to simulate grain-boundary motion, but to capture fully the interaction between this motion and other microscale phenomena that take place in pure polycrystalline materials, namely stress-mediated self-diffusion and electromigration. This distinguishes the current work from the existing literature.

In the current work, we recast the level-set equation in a simpler form by assuming that the level-set function remains a signed distance to the migrating grain boundary (as with the original level-set equation, the use of an extensional velocity field helps maintain this signed-distance function). \mbox{Mourad \etal\ \citep{Mourad:2005}} tested the resulting level-set formulation extensively and concluded that it is both accurate and robust despite its remarkable simplicity (for some interface-evolution problems, this approach reduces the original level-set equation, a nonlinear hyperbolic PDE, to an ODE that is almost trivial to solve). They conducted several numerical experiments to assess the ability of the simplified level-set scheme to capture the correct solution, particularly in the presence of discontinuities in the extensional velocity and/or in the gradient of the level-set function. They also examined the convergence properties of the method and its performance in a variety of problems, including curvature flow and problems where the simplified level-set equation takes the form of a Hamilton-Jacobi equation with convex or non-convex Hamiltonian. Discretizations based on structured and unstructured finite-element meshes of bilinear quadrilateral and linear triangular elements were shown to perform equally well. They also found that sufficient accuracy is available through a standard Galerkin formulation without resorting to any stabilization or discontinuity-capturing \citep{Barth:98,Hansbo:93} techniques.

In addition, a variant of the simplified level-set formulation mentioned above was employed by \mbox{Ji \etal\ \citep{Ji:2005}} for representing the evolution of phase boundaries over unstructured finite-element meshes. Here, we treat a complex coupled problem of which grain-boundary migration is only one facet, in addition to mechanics, self-diffusion and electromigration. We apply this simplified level-set scheme to the problem of grain-boundary migration within this context, and we demonstrate its implementation as an integral part of the wider computational framework used to solve the coupled problem under consideration.

The remainder of this paper is organized as follows. In Section~\ref{sec:coup}, we summarize the formulation for coupled mass transport and mechanics. Then we examine the thermodynamics and kinetics of grain-boundary migration, and show how this phenomenon interacts with mass transport and mechanics in polycrystals. In Section~\ref{sec:num}, we formulate the grain-boundary migration problem using the level set method and we describe the computational methods used in the implementation of this formulation. Numerical examples are presented in Section~\ref{sec:resu}. A summary is provided and conclusions are drawn in Section~\ref{sec:conc}.

\section{The coupled formulation} \label{sec:coup}

\subsection{Thermodynamic basis} \label{sec:coup_ther}

\begin{figure}[tb]
\begin{center}
\psfrag{TN}{$\bst=\bssigma\bsn$}
\psfrag{NN}{$\bsn$}
\includegraphics [width=2.5in] {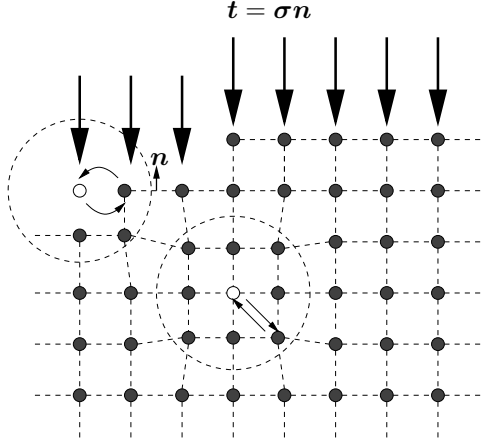}
\caption{Schematic rendering of a lattice with atoms and vacancies.}
\label{fig:lattice}
\end{center}
\end{figure}

The thermodynamics is posed in a continuum setting, with motivation provided by atomic processes. A schematic of the lattice is shown in Fig.~\ref{fig:lattice}. It shows atoms, vacancies, and a free surface; the latter is a source and sink for vacancies. A grain boundary could serve as a source or sink also. Free surfaces and grain boundaries are treated as regions of finite width, $\delta_{s}$ and $2\delta_{gb}$ respectively.

\subsubsection{Internal energy density}

We consider crystalline materials in which the dangling bonds around a vacancy cause an inward relaxation of the surrounding lattice (see Fig.~\ref{fig:lattice}). The resulting vacancy relaxation strain can be expressed as
\begin{equation} \label{eq:epv}
\bsepsilon^{v}=-\dfrac{1}{3}(1-f)\Omega(\conc-\ceqo)\bsone,
\end{equation}
where $\Omega$ is the atomic volume, $f\Omega$ (with $0<f<1$) is the volume of a vacancy, $\conc$ is the vacancy concentration, $\ceqo$ is the vacancy concentration at thermodynamic equilibrium under vanishing external stress and $\bsone$ is the second-order isotropic tensor. The creep strain resulting from the accumulation or depletion of atoms at a free surface or grain boundary with unit normal, $\bsn$, can be expressed as
\begin{equation}
\bsepsilon^{c}=\dfrac{1}{3}\theta_{c}(\bsn\otimes\bsn).
\end{equation}
A non-phenomenological evolution equation for $\theta_{c}$ was derived and discussed in detail by \mbox{Garikipati \etal\ \citep{Garikipatietal:01}}. The thermal strain is given by
\begin{equation}
\bsepsilon^{th}=\alpha(T-T_{0})\bsone,
\end{equation}
where $T$ is the temperature, $T_{0}$ is a reference temperature and $\alpha$ is the linear coefficient of thermal expansion. The elastic strain is obtained by subtracting these inelastic strain contributions from the total strain, $\bsepsilon$; i.e.
\begin{equation} \label{eq:epe}
\bsepsilon^{e}=\bsepsilon-(\bsepsilon^{v}+\bsepsilon^{c}+\bsepsilon^{th}).
\end{equation}
The stress is obtained from the elastic strain and the (generally anisotropic) fourth-order elasticity tensor, $\elasc$, as
\begin{equation} \label{eq:str}
\bssigma=\elasc:\bsepsilon^{e}.
\end{equation}
Given a state with local stress, $\bssigma$, the incremental elastic strain energy density is then given as $\bssigma:\delta\bsepsilon^{e}=(\elasc:\bsepsilon^{e}):\delta\bsepsilon^{e}$, where $\delta\bsepsilon^{e}$ is the increment in elastic strain. With this background, the incremental internal energy density $\delta e$, corresponding to a state $\{\bsepsilon,T,\conc\}$, and increments $\delta\bsepsilon$ and $\delta\conc$ is
\begin{equation} \label{eq:ine}
\delta e=\delta\hat{e}_{\eta}(\eta)+\evf\delta\conc+(\elasc:\bsepsilon^{e}):\delta\bsepsilon^{e},
\end{equation}
where $\eta$ is the entropy density and $\evf$ is the vacancy formation energy. The specific form of $\hat{e}_{\eta}(\eta)$, the entropic dependence of $e$, is not important to the development that follows.

\subsubsection{External work density}

The density of work done by external agents can be expressed as
\begin{equation}
\delta w_{ext}=\bssigma:\delta\bsepsilon-q\psi\delta\conc+\dfrac{3}{2}(\bsn\cdot\bssigma\bsn)f\Omega\delta\conc\chi.
\end{equation}
Here, the first term of the right-hand side is the classical stress-power term. The second term accounts for the apparent work performed by the electrostatic potential, $\psi$, during electromigration; in this phenomenological treatment, $q$ is the apparent charge ascribed to each vacancy. The last term accounts for the work done against the stress when vacancies are created at a free surface or grain boundary. The numerical factor appearing in this term arises from geometrical considerations. Additionally, the requirement that this term be active only at sources and sinks is enforced using the indicator function, $\chi$, defined as
\begin{equation} \label{eq:chi}
\chi(\bsx,t)=\left\{\begin{array}{rl}
1\quad&\text{if}\:\bsx\:\text{is in a surface- or grain-boundary region},\\
0\quad&\text{otherwise}.
\end{array}\right.
\end{equation}

\subsubsection{Entropy density}

The total entropy density is given by
\begin{equation}
\eta=\hat{\eta}_{vib}(T)-k\left[\conc\log\left(\dfrac{\conc}{\cs}\right)+\ca\log\left(\dfrac{\ca}{\cs}\right)\right],
\end{equation}
Since the formulation is isothermal, the specific form of the vibrational term, $\hat{\eta}_{vib}(T)$, is unimportant. The second term is the entropy density due to mixing (see Kittel and Kroemer~\citep{KittelKroemer:80} for details), $k$ is the Boltzmann constant, $\ca$ is the concentration of atoms, and $\cs$ is the lattice site concentration. Assuming that $\cs$ remains fixed in any material volume, i.e.\ $\delta\cs=\delta\ca+\delta\conc=0$, the incremental entropy density corresponding to an increment in vacancy concentration, $\delta\conc$, at given stress and temperature can be expressed as
\begin{equation}
\delta\eta=-k\log\left(\dfrac{\conc}{\ca}\right)\delta\conc.
\end{equation}

\subsubsection{The Gibbs free energy density}

At the given state of stress and temperature, the incremental Gibbs free energy density corresponding to increments $\delta\bsepsilon$ and $\delta\conc$ is defined as
\begin{equation} \label{eq:gib}
\delta g=\delta e-\delta w_{ext}-T\delta\eta.
\end{equation}
With this, the constitutive relations can be obtained in a systematic fashion as outlined in the following section.

\remark Since free surfaces and grain boundaries are considered, there are accompanying surface and grain boundary energies, $\gamma_{s}$ and $\gamma_{gb}$. These energies are taken to be independent of the strain and vacancy concentration for this formulation, and therefore do not appear in the incremental Gibbs free energy density.

\subsection{Constitutive relations} \label{sec:coup_crel}

Anisotropic elasticity is assumed and the relation between the stress, $\bssigma$, and the elastic strain, $\bsepsilon^{e}$, is given by (\ref{eq:str}). The relation between the current density, $\bsi$, and the electric potential, $\psi$, is given by Ohm's law:
\begin{equation}
\bsi=-\dfrac{\bsnabla\psi}{\rho},
\end{equation}
where $\rho$ is the electric resistivity.

\subsubsection{The chemical potential of vacancies}

The vacancy chemical potential is defined in the usual fashion~\citep{Gibbs:28}:
\begin{equation} \label{eq:muv}
\muv\delta\conc=\muvo\delta\conc+\delta g,
\end{equation}
where $\muvo$ is a constant reference potential. Applying (\ref{eq:muv}) to (\ref{eq:ine}--\ref{eq:gib}) gives
\begin{equation} \label{eq:muv2}
\muv=\muvo+\evf+\left(\elasc:\bsepsilon^{e}\right):\dfrac{1}{3}(1-f)\Omega\bsone-\dfrac{3}{2}(\bsn\cdot\bssigma\bsn)f\Omega\chi+q\psi+kT\log\left(\dfrac{\conc}{\ca}\right).
\end{equation}
The coupling with mechanics is evident through the strain- and stress-dependent terms.

\subsubsection{The vacancy flux}

The vacancy flux is obtained from (\ref{eq:muv2}) via the relation
\begin{equation} \label{eq:flux0}
\bsj_{v}=-\dfrac{D_{v}\conc}{kT}\bsnabla\muv,
\end{equation}
where $D_{v}$ is the vacancy diffusivity. This relation has been derived from an atomic basis by Bardeen~\citep{Bardeen:49}. The term $D_{v}\conc/kT$ is the mobility of a vacancy and $-\bsnabla\muv$, the force acting on it, is the driving force for mass transport.

\subsection{Governing equations} \label{sec:coup_PDEs}

The constitutive relations established above are incorporated in balance laws for mechanics, electric flow and mass transport, leading to a coupled system of governing differential equations.

\subsubsection{Mechanics}

Neglecting dynamic effects and body forces, the mechanics problem is governed by the quasistatic equilibrium equation and appropriate boundary conditions:
\begin{subequations}
\begin{align}
\bsnabla\cdot\bssigma&=\boldsymbol{0},&&\text{in}\:\msb,\\
\bsu&=\bar{\bsu},&&\text{on}\:\partial\msb_{\bsu},\\
\bssigma\bsn&=\bar{\bst},&&\text{on}\:\partial\msb_{\bssigma},
\end{align}
\end{subequations}
where $\msb$ is the domain of interest and the boundary subsets $\partial\msb_{\bsu}$ and $\partial\msb_{\bssigma}$ have essential and natural boundary conditions specified, respectively. These subsets satisfy $\partial\msb_{\bsu}\cap\partial\msb_{\bssigma}=\varnothing$ and $\overline{\partial\msb_{\bsu}\cup\partial\msb_{\bssigma}}=\partial\msb$.

\subsubsection{Electric flow}

The electric flow problem is governed by the charge conservation equation with Dirichlet and Neumann boundary conditions:
\begin{subequations}
\begin{align}
\bsnabla\cdot\bsi&=0,&&\text{in}\:\msb,\\
\psi&=\bar{\psi},&&\text{on}\:\partial\msb_{\psi},\\
\bsi\cdot\bsn&=\bar{\imath},&&\text{on}\:\partial\msb_{\bsi},
\end{align}
\end{subequations}
where $\partial\msb_{\psi}\cap\partial\msb_{\bsi}=\varnothing$ and $\overline{\partial\msb_{\psi}\cup\partial\msb_{\bsi}}=\partial\msb$.

\subsubsection{Mass Transport}

The mass-transport problem is governed by the continuity equation for vacancies, and appropriate initial and boundary conditions:
\begin{subequations}
\begin{alignat}{3}
\dfrac{\partial\conc}{\partial t}&=-\bsnabla\cdot\bsj_{v}-\dfrac{1}{\tau}(\conc-\ceq)\chi,&\qquad&\text{in}\:\msb,&\:&\:t\geq0,\\
\conc&=\conc^{0},&\qquad&\text{in}\:\msb,&\:&\:t=0,\\
\conc&=\bar{C}_{v},&\qquad&\text{on}\:\partial\msb_{\conc},&\:&\:t\geq0,\\
\bsj_{v}\cdot\bsn&=\bar{j},&\qquad&\text{on}\:\partial\msb_{\bsj_{v}},&\:&\:t\geq0,
\end{alignat}
\end{subequations}
where $\partial\msb_{\conc}$ and $\partial\msb_{\bsj_{v}}$ are the boundary subsets on which concentration and flux boundary conditions are specified, respectively. Here, the boundary subsets satisfy $\partial\msb_{\conc}\cap\partial\msb_{\bsj_{v}}=\varnothing$ and $\overline{\partial\msb_{\conc}\cup\partial\msb_{\bsj_{v}}}=\partial\msb$. The effectiveness of vacancy sources and sinks is characterized by the relaxation time, $\tau$. The equilibrium vacancy concentration, $\ceq$, is defined by $\muv|_{\ceq}=\muvo$ in~(\ref{eq:muv2}).

\subsection{Grain-boundary migration} \label{sec:GBM}

\begin{figure}[tb]
\begin{center}
\psfrag{e}[Br]{$\evf$, $\evd$}
\psfrag{Grain1}[]{Grain A}
\psfrag{Grain2}[]{Grain B}
\psfrag{GBR}[]{\scriptsize{Grain-boundary region}}
\psfrag{SBR}[]{\scriptsize{Surface-boundary region}}
\psfrag{g}[]{\footnotesize{2$\delta_{gb}$}}
\psfrag{s}[]{\footnotesize{$\delta_{s}$}}
\includegraphics [width=3.0in] {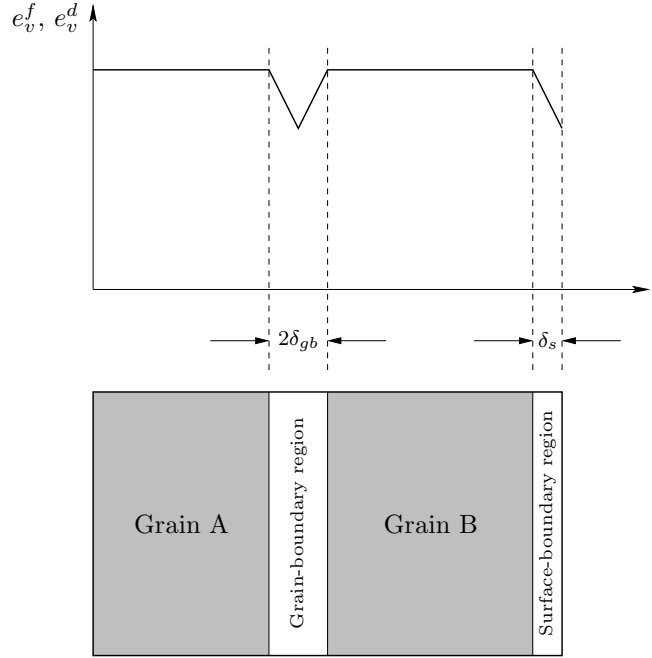}
\caption{Variation of $\evf$ and $\evd$ across boundary regions.}
\label{fig:interfaces}
\end{center}
\end{figure}

The current location of a migrating grain boundary determines the value of the indicator function, $\chi(\bsx,t)$ (see Eq.~(\ref{eq:chi})); i.e.\ it determines whether vacancy sources/sinks are active and whether creep strain can accumulate at a given point, $\bsx\in\msb$. Furthermore, information about the location of the boundary is needed to calculate the value of the vacancy formation energy, $\evf$, and the activation energy for diffusion, $\evd$, everywhere in the domain of interest. These properties are assumed to vary linearly over the width of a boundary region as shown in Fig.~\ref{fig:interfaces}.

The foregoing illustrates the influence of grain-boundary migration on mass transport. Due to the tight coupling between mass transport and mechanics, the migration of the grain boundary also affects the stress. In turn, mass transfer across the grain boundary causes one grain to grow at the expense of its neighbor and thus leads to grain-boundary migration.

\subsubsection{Thermodynamic driving forces} \label{sec:migr_theo_ther}

Generally, interface migration in polycrystals is driven by the accompanying decrease in the free energy of the system. The thermodynamic driving force for such a process, acting on a unit area of the interface, is thus defined as
\begin{equation} \label{eq:m_dforc}
p=-\dfrac{\delta\mcg}{\delta V},
\end{equation}
where $\delta\mcg$ is the increase in the total Gibbs free energy of the system brought about by a motion of the interface, during which the interface sweeps through the volume $\delta V$. Neglecting triple junctions, the Gibbs free energy of a polycrystal can be expressed as the sum of two contributions:
\begin{equation} \label{eq:m_gfe}
\mcg=\int\limits_{\msb}g~\rmda+\displaystyle\sum\limits_{i}\int\limits_{\msf_{i}}\gamma_{i}~\rmdl,
\end{equation}
where $g$ is the Gibbs free energy density as defined in (\ref{eq:gib}) and $\gamma_{i}$ is the energy per unit area of an interface, $\msf_{i}$. The term \emph{interface} is used here to refer to free surfaces as well as grain boundaries, and the summation in (\ref{eq:m_gfe}) is over all such interfaces in the polycrystal. 

In situations where the Gibbs free energy density, $g$, suffers a decrease across a grain boundary, the total free energy of the system can be reduced if the grain with the smaller value of $g$ (evaluated at the grain boundary) were to grow at the expense of its neighbor. Thus, a driving force acts on the grain boundary. For example, during recrystallization, annealed grains grow at the expense of cold-worked grains in which large dislocation densities lead to high values of $g$. The misorientation between two adjacent grains of an elastically anisotropic material subjected to a directional load causes one grain to store a smaller amount of strain energy per unit volume than its neighbor. This leads to strain-induced grain-boundary migration. Electromigration also causes atoms to jump across grain boundaries and thus leads to the migration of these boundaries~\citep{Katsman:97}.

Under the current formulation, the driving force discussed above can be characterized as follows: When a single atom hops across the grain boundary, it exchanges positions with a vacancy. Therefore, the volume change associated with this hopping event is \mbox{$\delta V=(1-f)\Omega$}. The decrease in the Gibbs free energy of the system is equal to the difference in the chemical potential of atoms across the boundary; i.e.\ \mbox{$-\delta\mcg=\Delta\mu_{a}$}, or equivalently \mbox{$-\delta\mcg=-\Delta\muv$}. Hence, from Eq.~(\ref{eq:m_dforc}), and assuming that the chemical potential gradient across the boundary is essentially linear~\citep{Turnbull:51}, the driving force, $p_{g}$, can be expressed as
\begin{equation} \label{eq:m_pg}
p_{g}=\dfrac{-\Delta \muv}{(1-f)\Omega}\approx\dfrac{(-\bsnabla\muv\cdot\bsn)2\delta_{gb}}{(1-f)\Omega},
\end{equation}
where $\bsn$ is the unit normal to the grain boundary and $2\delta_{gb}$ is its width. 

\begin{figure}[tb]
\begin{center}
\psfrag{gammar}[Br]{$\gamma_{gb}$}
\psfrag{gamma}{$\gamma_{gb}$}
\psfrag{norm}{$\bsn$}
\psfrag{Rad}{$R$}
\psfrag{Press}{$\dfrac{\gamma_{gb}}{R}$}
\includegraphics [width=3.0in] {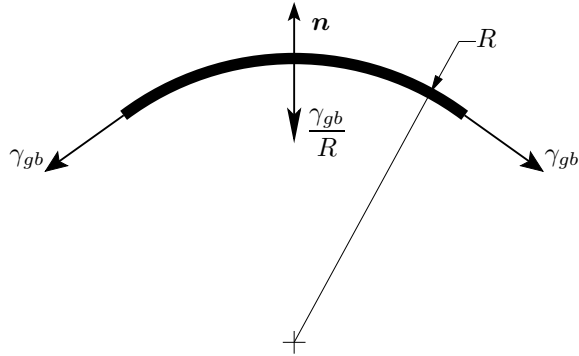}
\caption{Driving force on a cylindrical grain boundary with a radius of curvature, $R$, due to surface tension, $\gamma_{gb}$.}
\label{fig:m_pgamma}
\end{center}
\end{figure}

From Eq.~(\ref{eq:m_gfe}), it is clear that the total Gibbs free energy of a polycrystal can also be lowered by decreasing the total surface area of the grain boundaries in the system. When a curved interface moves away from its center of curvature, sweeping through an increment of volume $\delta V$, the free energy of the system increases by
\begin{equation} \label{eq:m_dgibbs}
\delta\mcg=\gamma_{i}\left(\dfrac{1}{R_{1}}+\dfrac{1}{R_{2}}\right)\delta V,
\end{equation}
where $\gamma_{i}$ is the (constant) specific surface energy of the interface and $R_{1}$ and $R_{2}$ are its principal radii of curvature. This expression can be traced back to Herring~\citep{Herring:51}. It follows that, in the current two-dimensional formulation, the driving force acting on a unit area of a (cylindrical) grain boundary, due to this effect, can be expressed as
\begin{equation} \label{eq:m_force2d}
p_{\gamma}=-\dfrac{\gamma_{gb}}{R},
\end{equation}
where $R$ is the radius of curvature and the negative sign indicates that $p_{\gamma}$ drives the boundary to migrate \emph{toward} its center of curvature (see Fig.~\ref{fig:m_pgamma}). In a polycrystal comprising only annealed grains, the action of $p_{\gamma}$ leads to normal grain growth; i.e.\ the growth of large grains at the expense of smaller ones. By contrast, in the early stages of recrystallization, small annealed grains grow at the expense of the surrounding matrix of cold-worked material. In this case, the boundaries surrounding the annealed grains move \emph{away} from their respective centers of curvature under the effect of $p_{g}$.

\subsubsection{Kinetic law} \label{sec:migr_theo_kine}

Assuming that grain-boundary migration takes place as a result of individual atoms hopping across the boundary, the migration velocity can be expressed in the following form:
\begin{equation} \label{eq:m_kine}
v_{n}=Mp,
\end{equation}
where $v_{n}$ is in the direction of the local unit normal, $\bsn$, which is assumed to point away from the boundary's center of curvature and $M$ is the grain-boundary mobility. This classical result, derived by Turnbull~\citep{Turnbull:51} using absolute reaction rate theory, holds provided that $p\Omega\ll kT$, a condition which is always met in grain growth and recrystallization~\citep{Gottstein:98}. The results of molecular dynamics simulations of curvature-driven~\citep{Upmanyu:98} and strain-induced~\citep{Schonfelder:97} grain-boundary migration in bicrystals agree with Eq.~(\ref{eq:m_kine}), which can be modified as follows, to account for both types of driving forces discussed in Section~\ref{sec:migr_theo_ther}:
\begin{equation} \label{eq:m_kine2}
v_{n}=M_{g}p_{g}+M_{\gamma}p_{\gamma}.
\end{equation}

\subsubsection{Grain-boundary mobility} \label{sec:migr_theo_mobi}

In Section~\ref{sec:migr_theo_ther}, it was established that diffusion of atoms across a grain boundary due to the local gradient in the atomic chemical potential causes the boundary to migrate. The driving force for boundary migration in this case, denoted $p_{g}$, is given by (\ref{eq:m_pg}). An expression for the corresponding mobility, $M_{g}$, can be obtained by stipulating that, in this case, the migration velocity in the direction of the local unit normal, $\bsn$, should be given by (see Porter and Easterling~\citep{PorterEasterling:2001})
\begin{equation} \label{eq:m_mg1}
M_{g}p_{g}=-\left(\bsj_{a}\cdot\bsn\right)\Omega-\left(\bsj_{v}\cdot\bsn\right)f\Omega,
\end{equation}
where $\bsj_{a}$ and $\bsj_{v}$ are, respectively, the local atomic and vacancy fluxes. Also recall that $\Omega$ is the atomic volume and $f\Omega$ is the volume of a vacancy. Since atoms and vacancies move by exchanging positions; i.e.\ $\bsj_{v}=-\bsj_{a}$, we have
\begin{equation} \label{eq:m_mg2}
M_{g}p_{g}=\left(\bsj_{v}\cdot\bsn\right)(1-f)\Omega.
\end{equation}
Finally, by combining (\ref{eq:m_mg2}), (\ref{eq:m_pg}) and (\ref{eq:flux0}), we obtain
\begin{equation} \label{eq:m_mg3}
M_{g}=\dfrac{D_{v}\conc(1-f)^{2}\Omega^{2}}{2\delta_{gb}kT}.
\end{equation}
Since grain-boundary migration under the sole influence of $p_{g}$ involves \emph{transport} across the boundary, and since, in the present treatment, mass transport is assumed to take place through the exchange of positions between atoms and vacancies, no lattice sites are transferred across the boundary in this process. By contrast, migration under the influence of $p_{\gamma}$ consists of the transfer of lattice sites across the grain boundary~\citep{Gottstein:98}, via the process of atoms detaching from one grain and attaching to the one on the opposite side of the boundary. It is hence unreasonable to expect the mobilities, $M_{g}$ and $M_{\gamma}$, of these two distinctly different processes to be the same.

As mentioned earlier, normal grain growth can be attributed to the action of the driving force $p_{\gamma}$. The fact that grain growth is a thermally activated process suggests an Arrhenius-type relationship between the mobility, $M_{\gamma}$, and the temperature:
\begin{equation} \label{eq:m_tempdep1}
M_{\gamma}=M_{0}\exp\left(\dfrac{-\egb}{kT}\right),
\end{equation}
where the pre-exponential factor, $M_{0}$, and the activation energy for grain-boundary migration, $\egb$, are dependent on the misorientation angle and the axis of rotation~\citep{Aust:59b,Upmanyu:99}.

Finally, by combining (\ref{eq:m_force2d}), (\ref{eq:m_kine2}), (\ref{eq:m_mg2}) and (\ref{eq:m_tempdep1}), we obtain the following expression for the migration velocity:
\begin{equation} \label{eq:m_kine3}
v_{n}=\left(\bsj_{v}\cdot\bsn\right)(1-f)\Omega-M_{0}\exp\left(\dfrac{-\egb}{kT}\right)\dfrac{\gamma_{gb}}{R}.
\end{equation}

\section{Computational methods} \label{sec:num}

In this section, we focus on the level-set formulation of the grain-boundary migration problem, and we present a detailed description of the computational techniques used in its implementation. To solve the coupled problem, the level-set formulation is integrated into the computational framework that was first introduced in Reference~\citep{Garikipatietal:01}. This computational framework is based on an operator-split solution scheme and relies on the finite element method to solve the mechanics, mass-transport and electric-flow problems individually. While Reference~\citep{Garikipatietal:01} addresses mainly the physics of stress-driven mass transport in polycrystals, it also contains details of the computational framework used therein. Here, these details are omitted; however, the operator-split solution scheme is outlined briefly in Section~\ref{sec:stagg} to show how the level-set formulation is incorporated into the computational framework used.

\subsection{Level-set formulation}

Evolving interfaces can be tracked using the level set method, originally introduced by Osher and Sethian~\citep{Osher:88}. A comprehensive review of the method and the computational algorithms used in its implementation can be found in~\citep{Sethian:99,Sethian:2001}. 

\begin{figure}[tb]
\begin{center}
\psfrag{R1}[]{$\msb^{-}$}
\psfrag{R2}[]{$\msb^{+}$}
\psfrag{H}{$\msf$}
\psfrag{dB}{$\partial\msb$}
\psfrag{ns}{$\bsn_{0}^{+}$}
\includegraphics [width=2.1in] {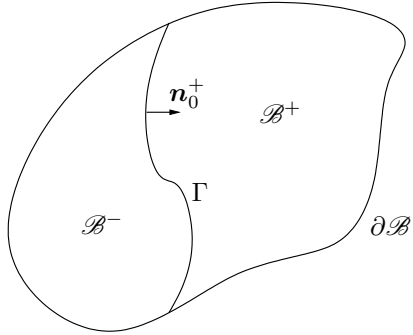}
\caption{Schematic of the domain of the grain-boundary migration problem.}
\label{fig:lsm}
\end{center}
\end{figure}

Consider an evolving grain boundary, $\msf$, which divides the domain of interest, $\msb$, into two disjoint open subsets, $\msb^{-}$ and $\msb^{+}$. This situation is depicted in Fig.~\ref{fig:lsm}. The boundary can be parameterized with the aid of the scalar function $\phi(\bsx,t)$, defined on $\msb$, provided that the following conditions are satisfied for all~$t\geq0$:
\begin{subequations} \label{eq:lscondi}
\begin{align}
\phi(\bsx,t)&<0,&&\forall\:\bsx\in\msb^{-},\\
\phi(\bsx,t)&=0,&&\forall\:\bsx\in\msf,\\
\phi(\bsx,t)&>0,&&\forall\:\bsx\in\msb^{+}.
\end{align}
\end{subequations}
The term \emph{level set} refers to a set of points with a fixed value of $\phi$, i.e.\ an iso-contour of $\phi$; the \emph{zero level set} represents the grain boundary. Accordingly, the unit normal to a given level set can be defined locally as
\begin{equation} \label{eq:lsnorm}
\bsn^{+}=\dfrac{\bsnabla\phi}{\|\bsnabla\phi\|},
\end{equation}
where $\|\cdot\|$ denotes the Euclidean norm. This expression can be evaluated at any point on the zero level set to obtain the local unit normal to the boundary, $\bsn_{0}^{+}$. This definition implies that $\bsn_{0}^{+}$ always points into the $\msb^{+}$ region.

The evolution of the level-set field is governed by
\begin{equation} \label{eq:lscont3}
\dfrac{\partial\phi}{\partial t}+F_{n}\,\|\bsnabla\phi\|=0,
\end{equation}
where $F_{n}(\bsx,t)$ is the (scalar) local propagation velocity of the level set passing through point $\bsx$. To track the motion of the grain boundary, we require that
\begin{equation} \label{eq:fnvn}
F_{n}(\bsx,t)=v_{n}(\bsx,t),\qquad\forall\:\bsx\in\msf,\;t\geq0,
\end{equation}
where $v_{n}$ is obtained at any point on the grain boundary from~(\ref{eq:m_kine3}). Although this requirement does not place any restrictions on the choice of $F_{n}$ away from $\msf$, the solution procedure is simplified greatly if $F_{n}$ is an \emph{extensional} velocity field; i.e.\ if
\begin{equation} \label{eq:lsext}
\bsnabla F_{n}\cdot\bsn^{+}=0.
\end{equation}
This first-order partial differential equation can be solved for $F_{n}$, at all points $\bsx\in(\msb\setminus\msf)$, using (\ref{eq:fnvn}) as a boundary condition. A simpler alternative strategy for constructing the extensional field is discussed in Section~\ref{sec:nume_leve_velo} below.

The level-set function, $\phi$, can be initialized as the signed distance from $\msf$ as follows:
\begin{equation} \label{eq:lsinit}
\phi(\bsx,0)=\left(\min_{\bsy\in\msf}\|\bsx-\bsy\|\right)\mathrm{sign}[(\bsx-\bsy)\cdot\bsn_{0}^{+}(\bsy)].
\end{equation}
It is noted that this initial condition satisfies (\ref{eq:lscondi}). It also implies that initially,
\begin{equation} \label{eq:lsncond}
\|\bsnabla\phi\|=1,\qquad\forall\bsx\in\msb.
\end{equation}
Importantly, this desirable mathematical property of the level-set field is preserved when the velocity field is extensional. This can be shown (see~\citep{Zhao:96}) by noting that
\begin{align}
\dfrac{\partial}{\partial t}\|\bsnabla\phi\|^{2}&=\dfrac{\partial}{\partial t}(\bsnabla\phi\cdot\bsnabla\phi) \nonumber\\
&=2\bsnabla\phi\cdot\dfrac{\partial}{\partial t}\bsnabla\phi \label{eq:proof1}.
\end{align}
Combining (\ref{eq:proof1}) and (\ref{eq:lscont3}) and assuming that $\phi$ and $F_{n}$ are smooth, we obtain
\begin{equation} \label{eq:proof2}
\dfrac{\partial}{\partial t}\|\bsnabla\phi\|^{2}=-2\bsnabla\phi\cdot\bsnabla F_{n}\|\bsnabla\phi\|-2\bsnabla\phi\cdot\bsnabla\|\bsnabla\phi\| F_{n}.
\end{equation}
Thus, if $F_{n}$ is extensional (i.e.\ $\bsnabla\phi\cdot\bsnabla F_{n}=0$) and $\phi$ is initially a signed-distance function (i.e.\ $\|\bsnabla\phi\|=1$ and $\bsnabla\|\bsnabla\phi\|=\boldsymbol{0}$), we have
\begin{equation} \label{eq:nochange}
\dfrac{\partial}{\partial t}\|\bsnabla\phi\|=0, 
\end{equation}
which implies that (\ref{eq:lsncond}) holds for all $t\geq0$. Hence, Eq.~(\ref{eq:lscont3}) reduces to
\begin{equation} \label{eq:lscont4}
\dfrac{\partial\phi}{\partial t}+F_{n}=0,
\end{equation}
which governs the evolution of $\phi(\bsx,t)$ from the initial condition (\ref{eq:lsinit}).

Recall that the mechanics, mass-transport and electric-flow problems are solved using the finite element method, and that part of the coupling between these three sub-problems, on one hand, and the grain-boundary migration sub-problem, on the other, is through the indicator function, $\chi(\bsx,t)$. In this setting, the level-set update is computed at each node via a generalized trapezoidal rule:
\begin{equation} \label{eq:lsupd}
\phi(\bsx_{A},t_{n+1})=\phi(\bsx_{A},t_{n})-\Delta t\, F_{n}(\bsx_{A},t_{n+a}),
\end{equation}
where $\bsx_{A}$ is the position vector of node $A$ and $t_{n+a}=at_{n+1}+(1-a)t_{n}$, with $0\leq a\leq 1$. It is noted that when explicit time integration is used ($a=0$), the update operation is trivial since $v_{n}$, and hence $F_{n}$, are determined from the known solution at $t=t_{n}$ and do not depend on $\phi(\bsx,t_{n+1})$. It is also noted that no stabilization is required due to the use of the simplified level-set equation (\ref{eq:lscont4}) in lieu of (\ref{eq:lscont3}). The advantages as well as the performance and stability characteristics of the resulting level-set formulation are studied in detail by \mbox{Mourad \etal\ \citep{Mourad:2005}}.

Additionally, $\chi$ is defined more precisely as follows:
\begin{equation} \label{eq:lschi}
\chi(\bsx,t)=H(\delta_{gb}-|\phi(\bsx,t)|),
\end{equation}
where $H(\cdot)$ is the Heaviside function.

\subsubsection{Velocity projection and field re-initialization} \label{sec:nume_leve_velo}

\begin{figure}[tb]
\begin{center}
\psfrag{H}{$\phi=0$}
\psfrag{ns}{$\bsn_{0}^{+}$}
\psfrag{n}{$\bsn^{+}$}
\psfrag{x}{$\bsx$}
\psfrag{bx}[Br]{$\bar{\bsy}$}
\psfrag{P}{$\phi=+c$}
\psfrag{Bp}{$\widetilde{\msb}^{+}$}
\psfrag{Bm}{$\widetilde{\msb}^{-}$}
\includegraphics [width=2.5in] {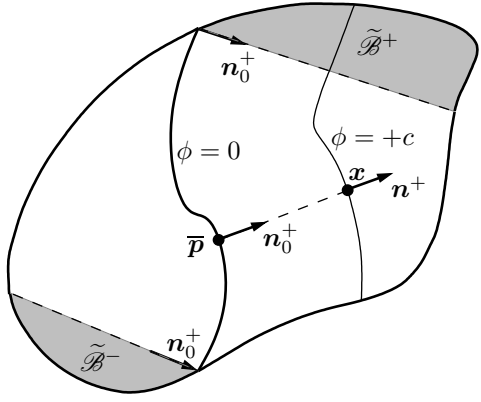}
\caption{Construction of the extensional velocity field by velocity projection.}
\label{fig:n_lsm3}
\end{center}
\end{figure}

The approach adopted here for constructing the extensional level-set propagation velocity field, $F_{n}$, is based on the notion that $F_{n}(\bsx)=v_{n}(\bar{\bsy})$ for any $\bsx\notin\msf$, if $\bar{\bsy}$ is such that
\begin{equation}
\|\bsx-\bar{\bsy}\|=\min_{\bsy\in\msf}\|\bsx-\bsy\|.
\end{equation}
It is clear that the resulting velocity field satisfies (\ref{eq:lsext}) when
\begin{equation}
\bsn_{0}^{+}(\bar{\bsy})=\dfrac{\bsx-\bar{\bsy}}{\|\bsx-\bar{\bsy}\|},
\end{equation}
as depicted in Fig.~\ref{fig:n_lsm3}. This is not always the case however; for instance, if $\bsx\in\widetilde{\msb}^{+}$ (see Fig.~\ref{fig:n_lsm3}), then $\bar{\bsy}$ is such that
\begin{equation}
\|\bsx-\bar{\bsy}\|=\min_{\bsy\in(\msf\cap\partial\msb)}\|\bsx-\bsy\|.
\end{equation}
It follows that, for all $\bsx\in\widetilde{\msb}^{+}$, $F_{n}(\bsx)=\text{const.}$, i.e.\ $\bsnabla F_{n}=\boldsymbol{0}$, which clearly satisfies (\ref{eq:lsext}) also. The above arguments apply in $\widetilde{\msb}^{-}$ as well. 

Due to the accumulation of numerical error, the level-set field may develop perturbations; i.e.\ $\left\|\bsnabla\phi\right\|$ may deviate from unity in some regions within $\msb$. The field must be re-initialized to neutralize these perturbations and retain accuracy by maintaining $\left\|\bsnabla\phi\right\|=1$ (see Eqs.~(\ref{eq:proof1}--\ref{eq:nochange})). Box~\ref{box:lsalgorithm} shows the algorithm used to implement the velocity projection scheme outlined above and the re-initialization scheme based on Eq.~(\ref{eq:lsinit}).

It is noted that, to preserve the integrity of the solution, the re-initialization procedure should not change the current location of $\msf$. In other words, the re-initialized field should have the same zero level set as the original (perturbed) field. The re-initialization scheme described here does not satisfy this condition strictly, i.e.\ it introduces error into the solution. Re-initialization should therefore be used judiciously. This issue is examined in detail by \mbox{Mourad \etal\ \citep{Mourad:2005}}. An alternative re-initialization procedure designed to minimize this type of error is described in detail in~\citep{Sussman:94,Sussman:99}. It must be noted however that, in some cases including curvature-driven migration, quadratic convergence in $L^{2}$ is achieved with the level-set update formula (\ref{eq:lscont4}), and importantly, this optimal convergence rate is preserved by the present re-initialization scheme (see~\citep{Mourad:2005,Peng:99} for details).

\begin{algorithm}[h]
\begin{center}
\fbox{\begin{minipage}{4.0in}
\texttt{FOR each node, $A$, DO} \\
\makebox[2em]{}\texttt{SET $\bsdd[A]=+\infty$} \\
\makebox[2em]{}\texttt{SET $\bsff[A]=0$} \\
\texttt{ENDDO} \\
\texttt{FOR each segment, $L_{0}$, of the zero level set DO} \\
\makebox[2em]{}\texttt{FOR each node, $A$, (with position vector $\bsx_{A}$) DO} \\
\makebox[4em]{}\texttt{FIND point $\bar{\bsz}\in L_{0}$ such that} \\
\makebox[5em]{}\texttt{$\displaystyle\|\bsx_{A}-\bar{\bsz}\|=\min_{\bsz\,\in\,L_{0}}\|\bsx_{A}-\bsz\|$} \\
\makebox[4em]{}\texttt{IF $\|\bsx_{A}-\bar{\bsz}\| < \left|\bsdd[A]\right|$ THEN} \\
\makebox[6em]{}\texttt{COMPUTE $v_{n}(\bar{\bsz})$} using Eq.~(\ref{eq:m_kine3}) \\
\makebox[6em]{}\texttt{COMPUTE $\bsn_{0}^{+}(\bar{\bsz})$} using Eq.~(\ref{eq:lsnorm}) \\
\makebox[6em]{}\texttt{SET $\bsff[A]=v_{n}(\bar{\bsz})$}\\
\makebox[6em]{}\texttt{SET $\bsdd[A]=\|\bsx_{A}-\bar{\bsz}\|\,\mathrm{sign}[(\bsx_{A}-\bar{\bsz})\cdot\bsn_{0}^{+}(\bar{\bsz})]$} \\
\makebox[4em]{}\texttt{ENDIF} \\
\makebox[2em]{}\texttt{ENDDO} \\
\texttt{ENDDO}\\
\texttt{FOR each node, $A$, DO} \\
\makebox[2em]{}\texttt{IF $\bsx_{A}\notin(\widetilde{\msb}^{+}\cup\widetilde{\msb}^{-})$ THEN} \\
\makebox[4em]{}\texttt{RE-INITIALIZE $\bsphi[A]=\bsdd[A]$} \\
\makebox[2em]{}\texttt{ENDIF} \\
\texttt{ENDDO}
\end{minipage}}
\caption{Velocity projection and level-set field re-initialization algorithm. Here, the global arrays, $\bsphi$, $\bsff$ and $\bsdd$ hold the nodal values of $\phi$, $F_{n}$ and $\|\bsx-\bsy\|$, respectively. A division of the zero level set which spans a single element is referred to as a \emph{segment} and is denoted by $L_{0}$. Also, since bilinear shape functions are used, an element can contain only one such segment.}
\label{box:lsalgorithm}
\end{center}
\end{algorithm}

The algorithm in Box~\ref{box:lsalgorithm} was first introduced by \mbox{Malladi \etal\ \citep{Malladi:95}} and was previously used by Garikipati and Rao~\citep{Garikipati:2001}. It is adopted here for its simplicity, despite being relatively expensive---the complexity of the present algorithm is at best $\order{n_{L}\times n_{np}}$, where $n_{L}$ is the number of elements intersected by the zero level set and $n_{np}$ is the total number of nodal points in the mesh (see~\citep{Mourad:2005}). A more efficient algorithm, such as the fast marching method~\citep{Sethian:99,Adalsteinsson:99} with $\order{n_{np}\log n_{np}}$ complexity, could be employed for velocity projection and level-set field re-initialization on Cartesian grids.

\subsubsection{Gradient smoothing} \label{sec:gradsmooth}

The local migration velocity, $v_{n}$, on the grain boundary is dependent on the local curvature, $\kappa=1/R$, which is defined as follows:
\begin{equation} \label{eq:n_ls_kappa0}
\kappa=\bsnabla\cdot\bsn^{+}.
\end{equation}
From (\ref{eq:lsnorm}) and (\ref{eq:lsncond}), it is clear that $\bsn^{+}=\bsnabla\phi$, and the curvature can hence be expressed as
\begin{equation} \label{eq:n_ls_kappa1}
\kappa=\nabla^{2}\phi=\dssum\limits_{i=1}^{n_{sd}}\dfrac{\partial^{2}\phi}{\partial x_{i}\partial x_{i}},
\end{equation}
where $n_{sd}=2$ is the number of spatial dimensions. Since the value of $\phi$ is updated at the finite element nodes, it is convenient to use the shape functions to evaluate the spatial derivatives in the above expression. However, since \emph{bilinear} shape functions are used for simplicity and robustness, Eq.~(\ref{eq:n_ls_kappa1}) cannot be used to evaluate $\kappa$ directly. To overcome this difficulty, we introduce a `smoothed' normal vector field $\tbsn^{+}$, weakly related to $\bsnabla\phi$ by
\begin{equation}
\int\limits_{\msb}\bsw\cdot(\tbsn^{+}-\bsnabla\phi)\,\rmda=0,
\end{equation}
where $\bsw$ is an arbitrary weighting function. Equivalently, the nodal values of each component, $\tn_{i}^{+}$, of the smoothed normal vector can be obtained by minimizing the following discretized functional:
\begin{equation}
\dssum\limits_{e=1}^{n_{el}}\int\limits_{\msb^{e}}\left[\dssum\limits_{A=1}^{n_{en}}\left(N_{A}\tn_{i}^{+}(\bsx_{A})-\dfrac{\rmd N_{A}}{\rmd x_{i}}\phi(\bsx_{A})\right)\right]^{2}\,\rmda,
\end{equation}
where $n_{el}$ is the number of elements in the model, $\msb^{e}$ denotes an element domain, $n_{en}$ is the number of nodes per element and $N_{A}$ is the shape function associated with node $A$. This leads to a matrix equation of the form $\bsmm\bsd=\bsf$, where $\bsd$ is the global vector containing the nodal values of the component $\tn_{i}^{+}$. The global mass matrix, $\bsmm$, and right-hand side vector, $\bsf$, are obtained from the corresponding element arrays via the usual assembly process. The element arrays in this case are given by
\begin{subequations} \label{eq:n_lssmoo_emat}
\begin{align}
m_{AB}^{e}&=\int\limits_{\msb^{e}}N_{A}N_{B}\,\rmda,\\
f_{A}^{e}&=\int\limits_{\msb^{e}}N_{A}\dssum\limits_{B=1}^{n_{en}}\dfrac{\rmd N_{B}}{\rmd x_{i}}\phi(\bsx_{B})\,\rmda.
\end{align}
\end{subequations}
Finally, the curvature is evaluated as follows:
\begin{equation}
\kappa=\dssum\limits_{A=1}^{n_{en}}\dssum\limits_{i=1}^{n_{sd}}\dfrac{\rmd N_{A}}{\rmd x_{i}}\,\tn_{i}^{+}(\bsx_{A}).
\end{equation}

\remark Using this least-squares technique to smooth the stress field leads to a mixed problem in stress and displacement form (see \nohyphens{Zienkiewicz and Taylor}~\citep{ZienkTay:2000}). By analogy, using this approach to compute the curvature is formally equivalent to a two-field mixed formulation for $\phi$ and $\tbsn^{+}$.

\remark Similar techniques have been used previously to evaluate the curvature of an evolving interface; e.g.\ see Chessa and Belytschko~\citep{Chessa:2003}.

\subsection{Operator-split algorithm} \label{sec:stagg}

The coupled problem is solved using an operator-split algorithm. The sequence of operations carried out in one time step is shown in Box~\ref{box:staggered} to illustrate how the level-set formulation is incorporated into this solution scheme. 

\begin{algorithm}[h]
\begin{center}
\fbox{\begin{minipage}{4.0in}
\makebox[2em]{(1)}\texttt{CONSTRUCT the smooth level-set gradient field, $\tbsn^{+}$.} \\
\makebox[2em]{(2)}\texttt{CONSTRUCT the extensional velocity field, $F_{n}$, and} \\
\makebox[2em]{}\texttt{RE-INITIALIZE the level-set field (see Box~\ref{box:lsalgorithm}).} \\
\makebox[2em]{(3)}\texttt{PERFORM the level-set update using Eq.~(\ref{eq:lsupd}).} \\
\makebox[2em]{(4)}\texttt{SOLVE the electric-flow problem for the electric} \\
\makebox[2em]{}\texttt{potential, $\psi$.} \\
\makebox[2em]{(5)}\texttt{REPEAT the following sequence:} \\
\makebox[5em][r]{(a)~~}\texttt{SOLVE the mechanics problem for the} \\
\makebox[5em]{}\texttt{displacements, $\bsu$.} \\
\makebox[5em][r]{(b)~~}\texttt{SOLVE the mass transport (composition)} \\
\makebox[5em]{}\texttt{problem for the vacancy concentration, $\conc$.} \\
\makebox[2em]{}\texttt{UNTIL both the mechanics and mass transport} \\
\makebox[2em]{}\texttt{problems have converged.} \\
\makebox[2em]{(6)}\texttt{INCREMENT time and GOTO step} (1).
\end{minipage}}
\caption{The operator-split algorithm used to solve the coupled problem.}
\label{box:staggered}
\end{center}
\end{algorithm}

\section{Numerical examples} \label{sec:resu}

\begin{table}[tb]
\caption{Material properties of Al (thin film) used in the analysis} \label{tab:v_mattab}
\begin{center}
\begin{tabular}{lcc}
\hline
\multicolumn{1}{c}{Parameter}                          & Value                  & Unit                             \\
\hline
\sm Elastic modulus, $c_{11}$                              & \sm $184.7$                & \sm GPa                              \\
\sm Elastic modulus, $c_{12}$                              & \sm $95.15$                & \sm GPa                              \\
\sm Elastic modulus, $c_{44}$                              & \sm $44.7$                 & \sm GPa                              \\
\sm Linear coefficient of thermal expansion, $\alpha$      & \sm $24\times10^{-6}$      & \sm $\mathrm{K}^{-1}$                \\
\sm Electric resistivity, $\rho$                           & \sm $4.2\times10^{-8}$     & \sm $\Omega.\mathrm{m}$              \\
\sm Apparent electric charge on a vacancy, $q$             & \sm $5.6077\times10^{-19}$ & \sm C                                \\
\sm Atomic volume in the absence of strain, $\Omega_{0}$   & \sm $16.61$                & \sm $\AA^{3}$                        \\
\sm Vacancy-atom volume ratio, $f$                         & \sm $0.8$                  &                                  \\
\sm Vacancy formation energy in the bulk, $\evf$           & \sm $0.67$                 & \sm eV                               \\
\sm Minimum value of $\evf$ in boundary regions${}^{\ast}$ & \sm $0.5433$               & \sm eV                               \\
\sm Activation energy for diffusion in the bulk, $\evd$    & \sm $1.47$                 & \sm eV                               \\
\sm Minimum value of $\evd$ in boundary regions${}^{\ast}$ & \sm $1.1090$               & \sm eV                               \\
\sm Grain-boundary width, $2\delta_{gb}$                   & \sm $0.198$                & \sm $\mathrm{\mu m}$                 \\
\sm Diffusivity premultiplier${}^{\dag}$, $D_{v_{0}}$      & \sm $2.6\times10^{3}$      & \sm $\mathrm{m}^{2}.\mathrm{s}^{-1}$ \\
\sm Reduced grain-boundary mobility premultiplier${}^{\ddag}$, $A_{0}$ & \sm $39.81$    & \sm $\mathrm{m}^{2}.\mathrm{s}^{-1}$ \\
\sm Activation energy for grain-boundary migration, $\egb$ & \sm $1.29$                 & \sm eV                               \\
\hline
${}^{\ast}$~~\footnotesize{See Fig.~\ref{fig:interfaces}.} && \\
${}^{\dag}$~~\footnotesize{The diffusivity is given by $D_{v}=D_{v_{0}}\exp(-\evd/kT).$} && \\
${}^{\ddag}$~~\footnotesize{$A_{0}=\gamma_{gb} M_{0}$; also see Eqs.~(\ref{eq:m_tempdep1}) and (\ref{eq:m_kine3}).} && \\
\hline
\end{tabular}
\end{center}
\end{table}

Numerical results, obtained by solving the coupled initial and boundary value problem, are presented here with the aim of highlighting some of the advantages of the current approach. A $1~\mathrm{\mu m}$ wide, $2.5~\mathrm{\mu m}$ long segment of an aluminum interconnect line is modeled. The values of the material parameters used for Al are given in Table~\ref{tab:v_mattab}. The segment consists of two pure Al crystals separated by a $\Sigma7$ tilt grain boundary ($38.2^{\circ}$ misorientation about $<$$111$$>$).

\begin{figure}[p]
\begin{center}
\psfrag{Vacancy Concentration}[]{\fontsize{8}{10}\selectfont Vacancy Concentration \normalsize}
\psfrag{unit}[]{\fontsize{8}{10}\selectfont [$\mathrm{cm^{-3}}$] \normalsize}
\includegraphics [width=4.5in] {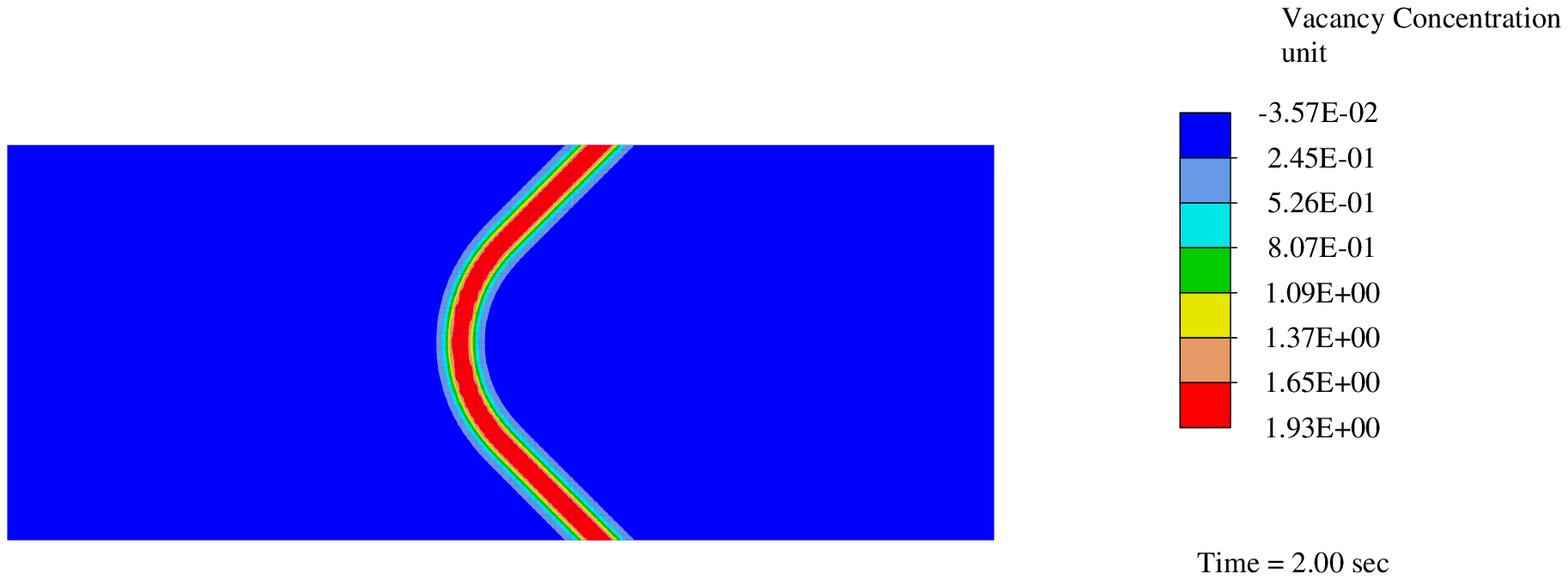}
\vskip 0.4in
\includegraphics [width=4.5in] {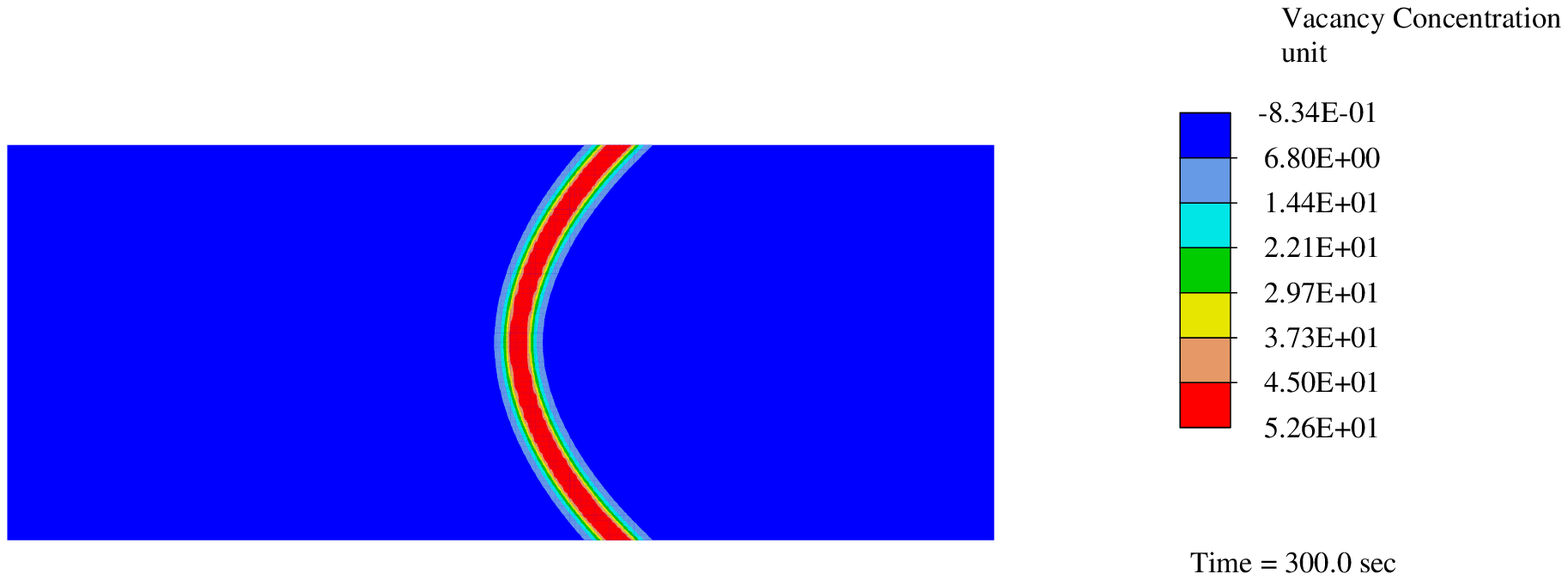}
\vskip 0.4in
\includegraphics [width=4.5in] {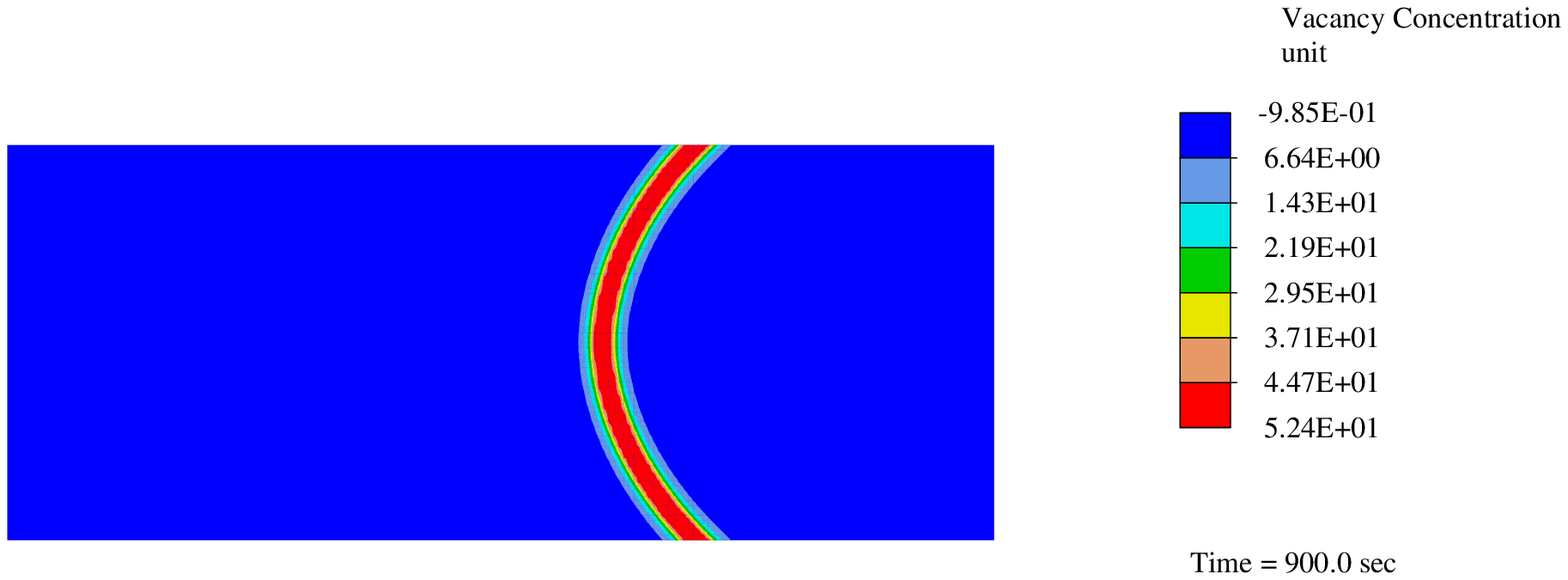}
\rput{0}(-12.4,14.12){(a)}
\rput{0}(-12.4,8.27){(b)}
\rput{0}(-12.4,2.40){(c)}
\caption{The evolution of the vacancy concentration contours in the interconnect line due to the motion of the grain boundary (first example). (a)~$t=2.0~\mathrm{sec}$. (b)~$t=300.0~\mathrm{sec}$. (c)~$t=900.0~\mathrm{sec}$.}
\label{fig:m_vacdist}
\end{center}
\end{figure}

The line is assumed to operate at $T=373~\mathrm{K}$, with a reference temperature, $T_{0}=473~\mathrm{K}$. Rigid passivation material surrounding the line prevents vacancies from crossing the upper and lower boundaries of the domain (see Fig.~\ref{fig:m_vacdist}a); i.e.\ the condition $\bsj_{v}\cdot\bsn=0$ holds at these boundaries. Periodic boundary conditions are imposed on the vacancy concentration at the left and right boundaries of the segment and an electrostatic potential difference, $\Delta\psi=0.0021~\mathrm{V}$, is applied between these two extremities. Vacancies drift along the electric field, $\boldsymbol{E}=-\bsnabla\psi$, pointing to the right.

In the first example, the migrating grain boundary consists, initially, of two straight (planar) sections, which are joined by a circular (cylindrical) section. The straight sections form $45^{\circ}$ angles with the upper and lower boundaries of the domain as shown in Fig.~\ref{fig:m_vacdist}a. Since the vacancy formation energy, $\evf$, is low inside grain-boundary regions where vacancy sources are also present, vacancies accumulate in such regions resulting in high values of the local vacancy concentration, $\conc$. The location of the grain boundary is revealed locally by the maximum-valued contour of $\conc$. 

Figure~\ref{fig:m_vacdist} shows the evolution of the vacancy concentration contours in the line. It is clear that, during the first $300$ seconds, the central cylindrical section of the boundary migrates to the right; i.e.\ toward its center of curvature, while the planar sections are relatively less mobile (Fig.~\ref{fig:m_vacdist}b). It must be emphasized that the interaction between mechanics, mass transport, electric effects and grain-boundary motion is accounted for, and the driving force for boundary migration, due to stress-driven diffusion and electromigration, is included in the calculations. However, its effect is overshadowed by the dominant driving force due to the curvature of the boundary. 

\begin{figure}[tb]
\begin{center}
\psfrag{Flux Magnitude}[]{\fontsize{8}{10}\selectfont Flux Magnitude \normalsize}
\psfrag{unit}[]{\fontsize{8}{10}\selectfont [$\mathrm{cm^{-2}.s^{-1}}$] \normalsize}
\includegraphics [width=4.5in] {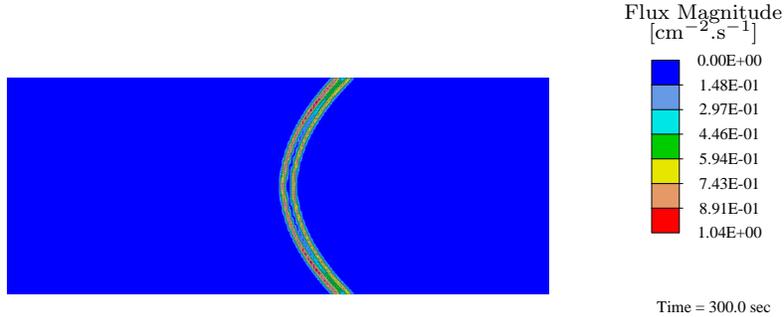}
\caption{Contours of the magnitude of the vacancy flux, $\|\bsj_{v}\|$, in the line (first example).}
\label{fig:m_fluxdist}
\end{center}
\end{figure}

\begin{figure}[tb]
\begin{center}
\includegraphics [width=3.2in] {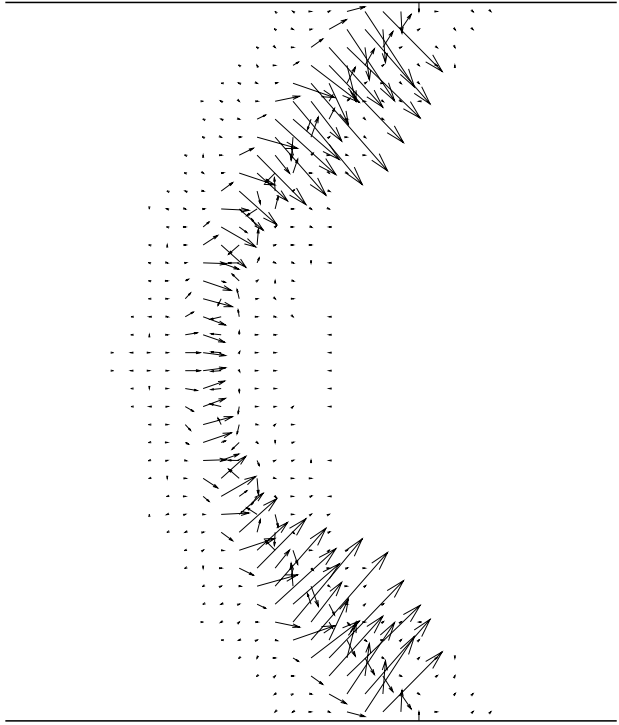}
\caption{Vector plot of the vacancy flux, $\bsj_{v}$, in the neighborhood of the grain boundary at $t=300~\mathrm{sec}$ (first example). Also see Fig.~\ref{fig:m_fluxdist} for magnitude of $\bsj_{v}$.}
\label{fig:m_vecflux}
\end{center}
\end{figure}

\begin{figure}[p]
\begin{center}
\psfrag{Vacancy Concentration}[]{\fontsize{8}{10}\selectfont Vacancy Concentration \normalsize}
\psfrag{unit}[]{\fontsize{8}{10}\selectfont [$\mathrm{cm^{-3}}$] \normalsize}
\includegraphics [width=4.5in] {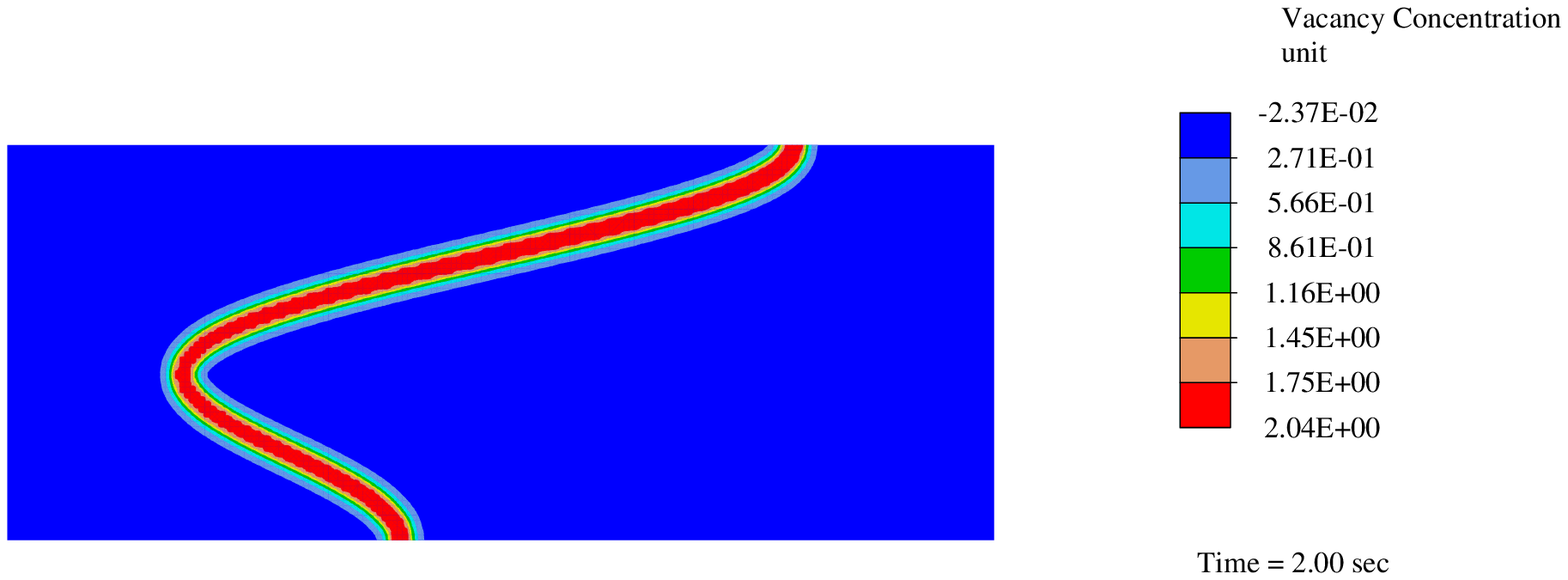}
\vskip 0.4in
\includegraphics [width=4.5in] {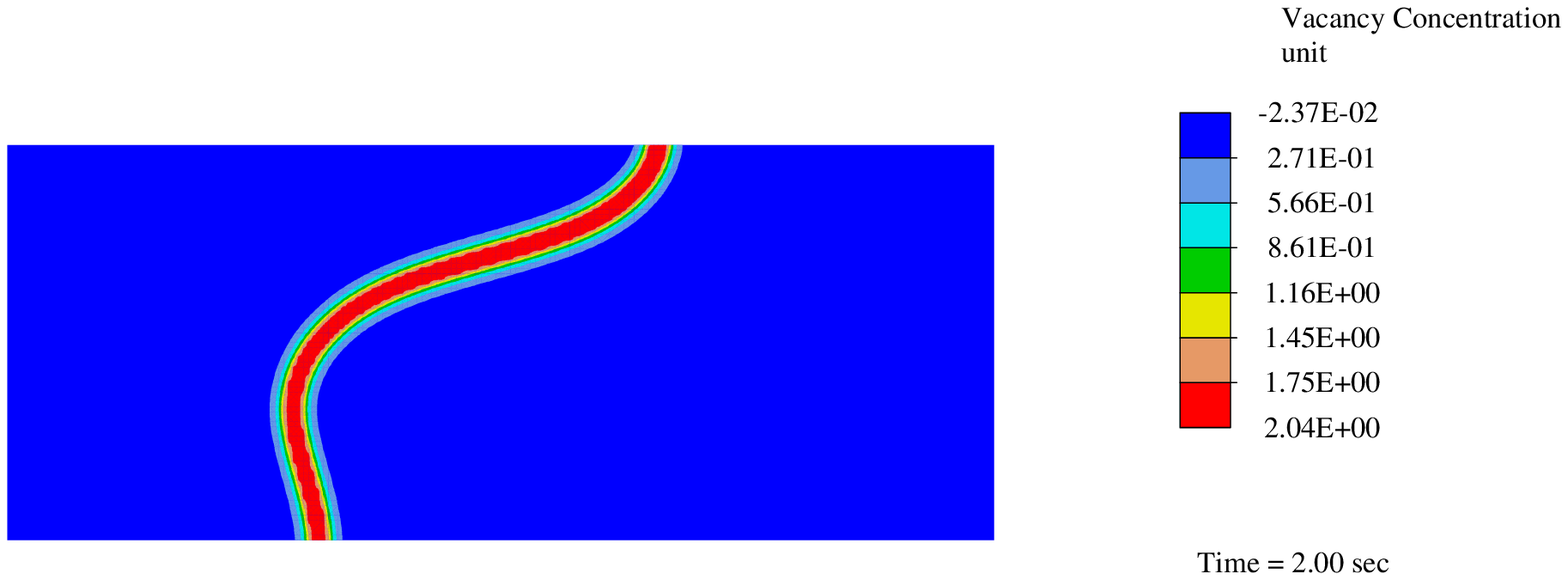}
\vskip 0.4in
\includegraphics [width=4.5in] {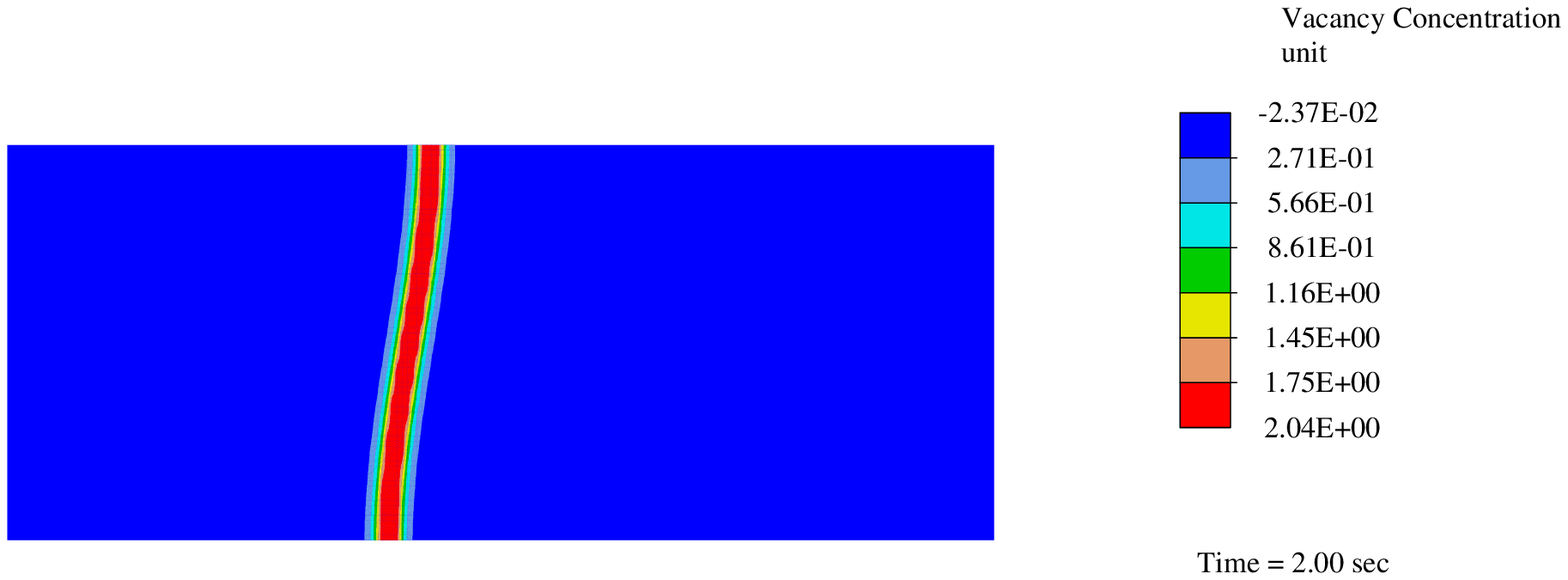}
\rput{0}(-12.4,14.12){(a)}
\rput{0}(-12.4,8.27){(b)}
\rput{0}(-12.4,2.40){(c)}
\caption{The evolution of the vacancy concentration contours in the interconnect line due to the motion of the grain boundary (second example). (a)~$t=2.0~\mathrm{sec}$. (b)~$t=200.0~\mathrm{sec}$. (c)~$t=1500.0~\mathrm{sec}$.}
\label{fig:m_vacdist2}
\end{center}
\end{figure}

After $900$ seconds, the curvature is approximately the same everywhere on the grain boundary (Fig.~\ref{fig:m_vacdist}c). At this stage, a steady state prevails and the boundary continues to travel toward the right without undergoing any further changes in shape. It is noted that the level-set calculations remain stable. The boundary remains smooth and does not develop any spurious cusps or ripples. The $45^{\circ}$ equilibrium angles between the grain boundary and the sidewalls are maintained as the solution progresses. 

A contour plot of the magnitude of the vacancy flux at $t=300~\mathrm{sec}$ is shown in Fig.~\ref{fig:m_fluxdist}. It is clear that a strong vacancy flux exists in the vicinity of the grain boundary. A vector plot of the vacancy-flux field in the neighborhood of the grain boundary is shown in Fig.~\ref{fig:m_vecflux} and although the field is complicated in this neighborhood, it can be seen that vacancies tend to drift in the same direction as the migrating boundary, thus preventing the appearance of a `trail' of vacancies or other oscillations in the numerical solution in the boundary's wake. It is important to note that the flux field evolves continually as the boundary migrates. It is also notable that the formulation captured this aspect of the coupling between mass transport and grain-boundary migration without additional terms being added to the expression of the vacancy flux to account, specifically, for the effect of moving boundaries.

The second example is concerned with the case where a curved grain boundary evolves into a planar configuration to reduce the free energy of the system. Here, the equilibrium angles at the grain boundary-sidewall intersections are set to $90^{\circ}$ and the initial geometry of the boundary is different and less regular than in the first example. It is clear from Fig.~\ref{fig:m_vacdist2}, which shows the evolution of the vacancy concentration contours in this case, that the boundary flattens as the solution progresses. It is noted that the equilibrium angles are also preserved in this case. It is also noted that in this case, perturbations in the solution lead to the formation of spurious `shadow' regions ($\widetilde{\msb}^{+}$, $\widetilde{\msb}^{-}$; see Fig.~\ref{fig:n_lsm3}) and re-initialization becomes necessary in these regions to maintain accuracy and to preserve the contact angles.

Although the numerical stability characteristics of the level-set formulation---and those of the operator-split algorithm---are not examined in a formal setting, no spatial or temporal oscillations are observed in the numerical solution of the example problems presented, as long as the time-step size is within the CFL limit; i.e.\ $\Delta t < h/F_{n}$, where $h$ is the mesh parameter. A detailed study of the performance and stability characteristics of the present level-set formulation is presented by \mbox{Mourad \etal\ \citep{Mourad:2005}}. Details regarding the advantages, applications and numerical stability characteristics of operator-split schemes can be found in~\citep{ArmSimo:92,ArmSimo:93,Felippa:2001} and references therein.

\section{Summary and Concluding Remarks} \label{sec:conc}

In this paper, we present the following contributions.

\begin{itemize}
\item \emph{\nohyphens{A computational formulation capable of capturing the full coupling between grain-boundary motion and other microscale phenomena that take place in pure polycrystalline materials}}. The coupled continuum formulation presented here was developed to model stress-driven self-diffusion and electromigration in polycrystals while accounting fully for the interaction between these mass transfer processes and the motion of grain boundaries. The formulation accounts for two distinct thermodynamic driving forces acting on a grain boundary; one due to the boundary's own curvature and another engendered by mass transfer across the boundary via stress-driven self-diffusion and electromigration. \\
\item \emph{\nohyphens{A simplified level-set formulation which does not require spatial stabilization}}. The level set method has \emph{previously} been used to pose grain-boundary migration as a time-dependent field problem governed by a pure advection equation. Standard numerical schemes resort to spatial stabilization techniques (e.g.\ upwinding schemes, Galerkin/Least-Squares) to attenuate the spurious oscillations known to appear in the numerical solution of equations of this type. Here, on the other hand, the level-set equation is reduced, using the mathematical properties of signed distance functions and extensional velocity fields, to a simpler form which obviates the need for these stabilization techniques. This leads to a remarkably simple explicit scheme for advancing the solution in time. The algorithm used to construct the extensional velocity field---and to simultaneously re-initialize the level-set field---is presented. We also provide the description of an $L^{2}$-projection technique used to compute the curvature of the grain boundary.
\end{itemize}

The numerical examples presented indicate that the strong coupling in the problem is captured adequately and that the numerical implementation allows the solution of the coupled initial and boundary value problem to be advanced in time in a stable fashion to obtain physically meaningful results. 

\section*{Acknowledgments}

We would like to acknowledge useful discussions with \mbox{Prof.\ G.\ M.\ Hulbert} of the University of Michigan. Support of this work by the National Science Foundation (grant CMS0075989) is gratefully acknowledged.

\end{document}